\documentclass[useAMS,usenatbib]{mn2e}
\usepackage{graphicx,url} \usepackage{multirow,amsmath} \usepackage{array}
\title{A new study of an old sink of sulfur in hot molecular cores: the sulfur
residue} \author[Paul~M. Woods et al.] {Paul M. Woods$^1$\thanks{E-mail:
p.woods@qub.ac.uk} A.~Occhiogrosso$^{1,2}$, S. Viti$^2$, Z. Ka\v{n}uchov\'a$^3$,
  M. E. Palumbo$^4$, \newauthor and S. D. Price$^5$\\
  $^1$Astrophysics Research Centre, School of Mathematics \& Physics, Queen's
  University, University Road, Belfast BT7 1NN, UK\\ $^2$Dept. of Physics \&
  Astronomy, University College London, Gower Place, London WC1E 6BT, UK \\
  $^3$Astronomical Institute of Slovak Academy of Science, T. Lomnica 05960,
  Slovakia\\ $^4$INAF - Osservatorio Astrofisico di Catania, via Santa Sofia 78,
  Catania 95123, Italy\\ $^5$Dept. of Chemistry, 20 Gordon Street, London WC1H 0AJ, UK\\
}
\begin{document}

\date{Accepted 2015 March 24}

\pagerange{\pageref{firstpage}--\pageref{lastpage}} \pubyear{2014}

\def\LaTeX{L\kern-.36em\raise.3ex\hbox{a}\kern-.15em
T\kern-.1667em\lower.7ex\hbox{E}\kern-.125emX}

\label{firstpage}

\maketitle

\begin{abstract} Sulfur appears to be depleted by an order of magnitude or more
from its elemental abundance in star-forming regions. In the last few years,
numerous observations and experiments have been performed in order to to understand
the reasons behind this depletion without providing a satisfactory explanation
of the sulfur chemistry towards high-mass star-forming cores. Several
sulfur-bearing molecules have been observed in these regions, and yet none are
abundant enough to make up the gas-phase deficit. Where, then, does this hidden
sulfur reside?  This paper represents a step forward in our understanding of the
interactions among the various S-bearing species. We have incorporated recent
experimental and theoretical data into a chemical model of a hot molecular core
in order to see whether they give any indication of the identity of the sulfur
sink in these dense regions. Despite our model producing reasonable agreement
with both solid-phase and gas-phase abundances of many sulfur-bearing species,
we find that the sulfur residue detected in recent experiments takes up only
$\sim$6 per cent of the available sulfur in our simulations, rather than
dominating the sulfur budget. \end{abstract}


\section{Introduction} Sulfur, despite being only the tenth most abundant
element in the Milky Way, is of significant astrochemical interest. Two of the
key open questions regarding sulfur are: {\it (i) where is the sulfur that
appears to be depleted from the gas phase in dense regions?} and, {\it (ii) what
use can studies of sulfur-bearing species be in our understanding of
astronomical environments?}

The depletion of sulfur became evident in the 1970s, 80s and 90s, when the
observed abundances of sulfur-bearing molecules in dense regions did not match
the observed cosmic abundance of sulfur
\citep[e.g.,][]{Penzias71,OppenheimerDalgarno74,Tieftrunk94,Palumbo97}, whereas
in diffuse and highly ionised regions, abundances of sulfur seemed roughly
cosmic \citep[$\sim$10$^{-5}$;
e.g.,][]{Savage96,Howk06,Martin-Hernandez02,Garcia-Rojas06}. This is often
referred to as the \textquoteleft sulfur depletion problem\textquoteright. Since
then, the interest in the chemistry of sulfur has heightened. To comprehend the
sulfur depletion problem fully it is important to go through the results that
have been published on this topic, in order to appreciate the state-of-the-art;
especially since relevant experimental work has been performed recently.

In 1999, it was proposed that sulfur typically existed in ionised form (S$^+$)
in translucent gas, and thus froze out more rapidly than neutrals upon cloud
collapse, due to electrostatic attraction to the negatively-charged grains
\citep{Ruffle99}. However, the form which sulfur takes upon freeze-out is not
evident: gas-phase carbonyl sulfide, OCS, has been detected in star-forming
regions \citep[SFRs;][]{vanderTak03}, at a level one thousand times too low
to be the dominant carrier of sulfur; but, estimates of grain-surface OCS
abundances may be affected by blending with a nearby methanol feature in
infrared spectra. Chemical models at the time suggested that S, SO, CS and
H$_2$S may be viable sinks of sulfur atoms
\citep[e.g.,][]{MillarHerbst90,Jansen95}, but our understanding of sulfur
chemistry, particularly the chemistry of CS, is not complete.

Our second key question asks about the intelligence that sulfur-bearing species
can give us in the understanding of astronomical environments. It has been
suggested by several authors that sulfur-bearing species may act as evolutionary
tracers for a specific region. In an attempt to study this potential role,
\citet{Charnley97} proposed that SO/H$_{2}$S and SO/SO$_{2}$ ratios act as
molecular clocks for grain mantle disruption since these ratios seem to vary
between different astronomical environments (dark clouds, hot cores, shocks and
winds around protostars were studied) and also within individual SFRs. 
\citet{Hatchell98} followed this idea, to constrain the ages of cores, by looking
at the ratios of sulfur-bearing species towards ultra-compact HII regions. In
particular, they developed different chemical models for each of the eight
sources they investigated, by varying different physical parameters.
\citet{Hatchell98} classified the sulfur-bearing species along with the age of
the young stellar object: \begin{itemize} \item H$_{2}$S, SO are abundant in
younger cores \item H$_{2}$S, SO, SO$_{2}$ at intermediate ages \item later, SO
and SO$_{2}$ are present, but without H$_{2}$S \item finally, CS, H$_{2}$CS and
OCS become the most abundant sulfur-bearing species. \citeauthor{Hatchell98}
suggested OCS is formed on the grain surface. \end{itemize}
A few years later, \citet{Viti01} proposed that NS/CS and SO/CS ratios were
specific indicators of a shock passage in the vicinity of a hot core.
In these physical conditions, the sulfur chemistry was found to be connected to
the HCO/H$_{2}$CO ratio. High values of these ratios indicated that a shock had
passed through the medium. 
Finally, \citet{Wakelam04} repeated the study by \citet{Hatchell98} and
highlighted how none of the ratios involving the four most abundant
sulfur-bearing species (H$_{2}$S, OCS, SO and SO$_{2}$) could be useful by itself
for estimating the core ages, because the amount of each molecule depends at
least on the physical conditions, the adopted grain mantle composition and also
evolutionary stage. 
A relatively recent paper by \citet*{Wakelam11} reported the study of S-bearing
species in four different high-mass dense core sources in order to investigate
the dependence of their abundances along with time. \citeauthor{Wakelam11} were
unable to reproduce the observed abundances for OCS, SO, SO$_{2}$, H$_{2}$S and
CS, but they found that the ratios between OCS/SO$_{2}$ and H$_{2}$S/SO$_{2}$
could be used to constrain some evolutionary time-scales. \citeauthor{Wakelam11}
also highlighted the difficulty in reproducing the amount of CS, which was
overestimated due to the fact that its abundance varies with radius and there is
an uncertainty about the location of the emitting region.

In order to fully understand what we know to date about the presence of
sulfur-bearing species in the interstellar medium, we present a summary of all
the species belonging to this family of molecules which have been observed
either in the gas-phase (Table~\ref{sulfurobserved}) or on the grain surface. In
the gas phase, sulfur is a ubiquitous element: it has been detected in different
astronomical environments, from the diffuse medium \citep{Liszt09} to dark
clouds \citep*{Dickens00}, as well as in hot cores and hot corinos
\citep{Schoier02, Sutton95}; in comets \citep{Boissier07}; in evolved stars
\citep{Woods03}; and in the atmosphere of Venus \citep{Krasnopolsky08}, in
various chemical forms. Its emission, therefore, has been widely observed, and
that has enabled the depletion of sulfur to be studied in a variety of
environments. For instance, \citet{Jenkins09} observed atomic sulfur lines
towards the diffuse medium and showed a relationship between the amount of
depletion of the elements and the density of the cloud. \begin{table*}
\begin{center} \caption{List of gas-phase sulfur-bearing species with the dense
sources towards which they were first observed. Example references are provided
in the third column. \label{sulfurobserved}} \begin{tabular}{clrl} \hline
Molecule & Source & Column density [cm$^{-2}$] & Reference\\ \hline CS & Orion
A, W51 & 2--210$\times$10$^{13}$ & \citet{Penzias71}\\ OCS & SgrB2 &
$\geq$3$\times$10$^{15}$ & \citet{Jefferts71}\\ H$_{2}$S & Hot cores &
4--50$\times$10$^{13}$ & \citet{Thaddeus72}\\ H$_{2}$CS & SgrB2 &
$>$1$\times$10$^{16}$ & \citet{Sinclair73}\\ SO & Orion A &
$\sim$1$\times$10$^{15}$ & \citet{Gottlieb73}\\ SO$_{2}$ & Orion A &
3--35$\times$10$^{15}$ & \citet{Snyder75}\\ SiS & IRC+10216, SgrB2 &
4$\times$10$^{13}$ & \citet{Morris75}\\ NS & SgrB2 & 1$\times$10$^{14}$ &
\citet{Gottlieb75,Kuiper75}\\ CH$_{3}$SH & SgrB2 & 2$\times$10$^{14}$ &
\citet*{Linke79}\\ HNCS & SgrB2 & 3$\times$10$^{13}$ & \citet*{Frerking79}\\
HCS$^{+}$ & SgrB2, Orion & 2--200$\times$10$^{11}$ & \citet*{Thaddeus81}\\
C$_{2}$S & TMC-1, SgrB2, IRC+10216 & 6--15$\times$10$^{13}$ &
\citet{Saito87,Cernicharo87}\\ C$_{3}$S & TMC-1, SgrB2 & 1$\times$10$^{13}$ &
\citet{Kaifu87,Yamamoto87}\\ SO$^{+}$ & IC 4434 & $\sim$5$\times$10$^{12}$ &
\citet{Turner92}\\ HSCN & SgrB2(N) & 1$\times$10$^{13}$ & \citet{Halfen09}\\
SH$^{+}$ & SgrB2 & $<$2$\times$10$^{14}$ & \citet{Menten11}\\ HS & W49N &
5$\times$10$^{12}$ & \citet{Neufeld12}\\ CH$_3$CH$_2$SH & Orion KL &
2$\times$10$^{15}$ & \citet{Kolesnikova14}\\ \hline \end{tabular} \end{center}
\end{table*}

On the other hand, the only two S-bearing species firmly detected on grain
surfaces have been OCS, with relatively low fractional abundances on the order
of 10$^{-7}$ \citep*{Palumbo95,Palumbo97} and SO$_{2}$
\citep{Boogert97,Zasowski09}, and thus the form of grain-surface sulfur is
relatively unknown. Several experiments were performed in order to understand
which molecules are candidates for explaining sulfur chemistry in the solid
state. In particular, \citet{Ferrante08} and \citet{Garozzo10} investigated the
mechanism of formation of OCS, discovering that CO reacts with free S atoms
produced by the fragmentation of the sulfur parent species. OCS was seen to be
readily formed by cosmic-ray irradiation, but at the same time it was easily
destroyed after continued exposure. \citet{Ferrante08} observed CS$_{2}$
production as one of the main product channels, although carbon disulfide has
not been yet detected in interstellar ices. An alternative molecule, hydrated
sulfuric acid, was suggested by \citet{Scappini03} as the main sulfur reservoir.
Later \citet*{Moore07} produced this species by ion irradiation of SO$_{2}$ and
H$_{2}$S in water-rich ice over the temperature range 86--130\,K.  More recent
papers by \citet{Garozzo10,Jimenez-Escobar11} and \citet*{Jimenez-Escobar14}
focused on the products of cosmic-ray and UV-photon irradiation of H$_{2}$S ice
analogues. Among the products, they found the presence of a sulfur residue which
might explain the missing sulfur in dense clouds. A revised chemistry of
S-bearing species has been discussed by \citet{Druard12}, who were able to
explain, in an initial attempt, the lack of observation of H$_{2}$S on the grain
surface, but without considering the presence of refractory sulfur.

The above summary of the state-of-art shows that, despite several
investigations, sulfur chemistry in the interstellar medium seems an intricate
problem that is yet to be solved. The present paper is a step forward in our
understanding of the sulfur chemistry in regions of star-formation. In
particular, we couple recent experimental data with a revised version of the
{\sc ucl\_chem} chemical model, taking into account the formation of this sulfur
residue. Moreover, we present a new classification of the thermal desorption of
sulfur species compared to the one elaborated by \citet{Viti04}. Non-thermal
desorption effects are also taken into account.  The paper is organised as
follows: Section~\ref{sec:mod} describes the astrochemical model;
Sect.~\ref{sec:expres} contains the details of all the experimental and
theoretical results that have been included in our chemical modelling, described
in two different subsections; Sect.~\ref{subsec:analtrends} contains the outputs
from {\sc ucl\_chem} models with their analyses and discussion;
Sect.~\ref{sec:obs} compares our theoretical results with data from the
observations; and finally, we present our conclusions in Sect.~\ref{sec:conc}.

\section{Modelling} \label{sec:mod}

We modify a pre-existing model, the {\sc ucl\_chem} chemical code
\citep{Viti99,Viti04}, in order to include the new experimental results.  Before
going into the detail of our updates, we briefly describe the physics and the
chemistry behind the model.  The model performs a two-step simulation.  Phase I
starts from a fairly diffuse medium where most chemical species are in atomic
form (apart from H$_{2}$), which undergoes a free-fall collapse until densities
typical of the gas that will form hot cores or hot corinos are reached (10$^{7}$
cm$^{-3}$ and 10$^{8}$ cm$^{-3}$, respectively). During this time, atoms and
molecules are depleted on to the grain surfaces and they hydrogenate when
possible. The depletion efficiency is determined by the fraction of the gas
phase material that is frozen on to the grains. This approach allows a
derivation of the ice composition by a time-dependent computation of the
chemical evolution of the gas-dust interaction process. The initial elemental
abundances of the main species (such as H, He, C, O, N, S and Mg) are the main
inputs for the chemistry, and are taken from \citet{Sofia01} and other
references detailed in \citet{Viti99}, in common with much recent {\sc
ucl\_chem} work. We usually assume that, at the beginning, only carbon and
sulfur are ionised and half of the hydrogen is in its molecular form. The other
elements are all neutral and atomic. Gas-phase reactions are taken from the
UMIST RATE06 database \citep{Woodall07}, and freeze-out reactions are included
in the reaction network in order to allow the formation of mantle species. The
depletion efficiency can be modified by adjusting the freeze-out fraction
parameter {\tt fr}. In Phase I we follow both the gas-phase and the
grain-surface chemistry, and the transitions between these two phases
(freeze-out and desorption).

Phase II is the warm-up phase and the model follows the gas-phase chemical
evolution of the remnant core when the protostar itself is formed. During
this stage, the importance of grain-surface reactions decreases with the
increasing temperature, because of the sublimation of important molecules (such
as CO) even at $\sim$20K \citep[see][]{Viti04}. The time-dependent evaporation
of the ice is treated in one of two ways. Before running the phase II model, we
can choose a final gas temperature ({\tt maxt}) for the astronomical object
studied.  The treatment of evaporation can be either time-dependent \citep[where
mantle species desorb in various temperature bands according to the experimental
results of][]{Collings04} or instantaneous (in that all species will desorb from
the grain surfaces at the first timestep). In this paper we only consider
time-dependent thermal desorption.  Non-thermal desorption of species is also
taken into account and is based on the study by \citet{Roberts07}. Three
desorption mechanisms are included in {\sc ucl\_chem}: desorption resulting from
H$_{2}$ formation on grains, direct cosmic-ray heating of the ice, and
cosmic-ray photodesorption. The latter mechanism is due to the generation of UV
photons which occur when cosmic-ray particles impact the grain. We do not
account for direct UV photodesorption, since the density, and hence UV
extinction, in these regions is generally high.

Finally, the outputs from the code consist of the fractional abundances of all
the species (in both gas phase and on the grain surface) as a function of time.
Fractional abundances of molecules are calculated as a ratio to the total number
of hydrogen nuclei ($n$(H) + $2n$(H$_{2}$)), where $n$ represents the number
density in cm$^{-3}$.

\section{Update of {\sc ucl\_chem} with recent results} \label{sec:expres}

\subsection{Experimental results}

We have updated the {\sc ucl\_chem} gas-grain chemical model to include some
published and unpublished experimental data on reactions involving S-bearing
species occurring on icy mantle analogues. Our aim is to derive more information
about the existence of a sulfur reservoir. First of all, we summarise the
results obtained from the different experiments (see
Subsections~\ref{subsec:h2scr},~\ref{subsec:labocs}) that we insert into the
chemical model. Then, we run models to benchmark the differences that the new
data make on the sulfur chemistry (see Sect.~\ref{sec:modgrid}).

\subsubsection{The effect of cosmic rays on icy H$_{2}$S mantle analogues}
\label{subsec:h2scr}

Cosmic rays impinging on icy mantles are energetic enough to induce chemical and
structural modifications on the grain surface. As a consequence the ice
constituents differ from the composition of the gas, and experiments of
astronomical relevance are therefore the only tools that can provide us with
information concerning the potential interactions that can arise after these
dynamic impacts.

\begin{figure}
\begin{center}
\includegraphics[width=90mm]{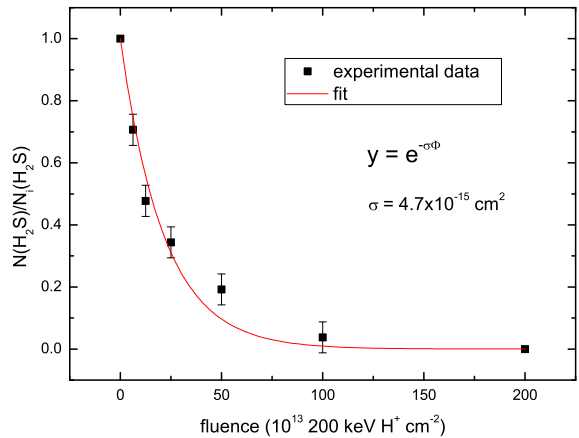} \caption{Column density ratio
N(H$_{2}$S)/N$_{i}$(H$_{2}$S) of hydrogen sulfide as a function of ion fluence
($\Phi$) after irradiation of the ice mixture. Data points are taken from
\citet{Garozzo10} and fitted with an exponential curve. \label{fig:expdata}}
\end{center} \end{figure}

In order to simulate a flux of cosmic-ray particles, \citet{Garozzo10}
irradiated their ice sample with 200 keV protons at 20 K in a high-vacuum
chamber (P $\le$ 10$^{-7}$\,mbar). Their sample consisted of a CO:H$_{2}$S=10:1
mixture. IR spectra were recorded before and after irradiation in order to
estimate the actual amount of each species produced from the ice mantle
analogues due to the proton bombardment. The integrated intensity measured for
each selected band (in optical depth $\tau$($\nu$) units) is proportional to the
column density of the species itself. Column densities for each species, $N$(X),
were calculated as the ratio with respect to the initial amount of H$_{2}$S,
N$_{i}$(H$_{2}$S), in the mixture. \begin{figure} \begin{center}
\includegraphics[width=90mm]{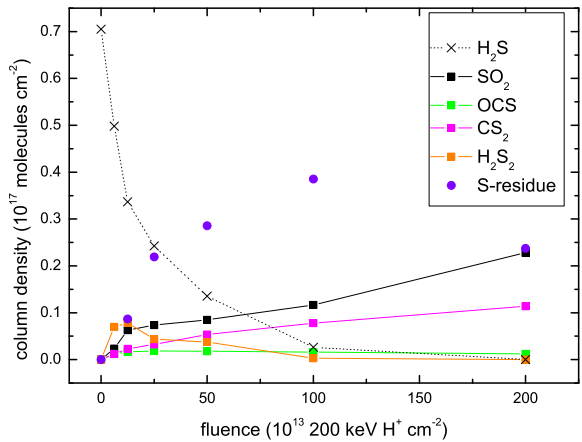} \caption{Column density of selected species as
a function of ion fluence after irradiation of a CO:H$_2$S = 10:1 ice mixture at
20\,K, using the data from \citet{Garozzo10}. Solid and dotted lines have been
drawn to guide the eye. \label{fig:ratiocolumn}} \end{center} \end{figure}
Figure~\ref{fig:ratiocolumn} shows the decrease of the initial amount of
hydrogen sulfide as the products of its dissociation form. The failure to detect
H$_{2}$S in the solid phase
\citep[\citealt*{Ehrenfreund04};][]{Garozzo10,Jimenez-Escobar11} may therefore
be linked to the strong ($\sim$ 80 per cent) reduction in its column density
when it is subjected to irradiation. This is in fact what was postulated by
\citet{Codella06} in order to justify the presence of a large amount of
gas-phase OCS observed in an extended, high velocity gas in the massive SFR, Cep
A East. We have analysed the experimental data collected by \citet{Garozzo10},
and in particular, we have fitted their data with an exponential curve (as
plotted in Fig.~\ref{fig:expdata}) in order to evaluate the reaction cross
section ($\sigma$= 4.7$\times$10$^{-15}$ cm$^{2}$). The identified molecules
with their specific production cross sections are listed in
Table~\ref{tab:detectmol}. The latter parameter was extrapolated by fitting the
experimental data at low fluence with a straight line.  \begin{table} \begin{center} \caption{A list of the
experimentally-detected sulfur-bearing species and their laboratory production
cross sections. \label{tab:detectmol}} \begin{tabular}{lc} \hline Molecule & $p$
[cm$^{2}$]\\ \hline H$_{2}$S$_{2}$ & 1.03$\times$10$^{-15}$\\ SO$_{2}$ &
6.78$\times$10$^{-16}$\\ CS$_{2}$ & 2.64$\times$10$^{-16}$\\ OCS &
2.05$\times$10$^{-16}$\\ S-residue & 1.19$\times$10$^{-15}$\\ \hline
\end{tabular} \end{center} \end{table}

Three important considerations come from the analysis of the data taken from
\citet{Garozzo10}: \begin{enumerate} \item A detection of CS$_{2}$ has not yet
been made in the ISM. Laboratory data allow us to quantify the abundance of this
molecule and to extend our model to include a reaction scheme for its formation
and loss channels.\\ \item There is evidence of a residue of species containing
sulfur, that can be estimated as follows: \begin{multline} \label{eq:residue}
N(\text{S-residue})=N_{i}(\mathrm{H_{2}S})-N(\mathrm{H_{2}S})-N(\mathrm{SO_{2}})
-N(\mathrm{OCS}) \\ \quad -2N(\mathrm{CS_{2}})-2N(\mathrm{H_{2}S_{2}})
\end{multline} This residue provides a constant supply of sulfur and it may
affect the role of sulfur as an evolutionary tracer. Moreover, as stated by
\citet{Anderson13}, the sputtering of this residue from the surface could lead
to the release of the large amount of atomic sulfur seen in shock regions.\\
\item These experiments help to predict the different forms that sulfur can take
on the grain surface and to estimate the ratios among sulfur-bearing molecules.
\end{enumerate}

We therefore insert into {\sc ucl\_chem} the chemical reactions experimentally
investigated by \citet{Garozzo10} and reported in Table~\ref{tab:dissoh2s}. 
Note that the reactions listed in Table~\ref{tab:dissoh2s} are simplified
representations of a complex experimental process. The action of the high-energy
protons on the laboratory ice analogues causes $\sim$10$^5$ molecular bonds to
break, leading to a chain of recombination reactions. As a result of these very
rapid recombinations, the products on the right-hand side are observed. These
reactions do not take place in thermodynamic equilibrium. The remaining sulfur, or
mSres, stays as a refractory element on the surface.  

\begin{table} \begin{center} \caption{Dissociation of solid H$_{2}$S due to
cosmic-ray impact (CR) experimentally investigated by \citet{Garozzo10}. All the
reaction channels are provided with a rate (in s$^{-1}$) of ISM relevance. The
$m$ before the molecular formula stands for {\it mantle}. \label{tab:dissoh2s}}
\begin{tabular}{lc} \hline Reaction & $k = \alpha$ [s$^{-1}$]\\ \hline
\phantom{2}mH$_{2}$S + \phantom{2m}CR$_{\phantom{1}}$ $\rightarrow$ mSres +
H$_2$ & 9.53$\times$10$^{-17}$\\ \phantom{2}mH$_{2}$S +
\phantom{2}mCO$_{\phantom{1}}$ $\rightarrow$ mOCS + H$_2$ &
1.65$\times$10$^{-17}$\\ \phantom{2}mH$_{2}$S + \phantom{2}mH$_{2}$S\phantom{\,}
$\rightarrow$ mH$_{2}$S$_{2}$ + H$_2$ & 8.23$\times$10$^{-17}$\\ 2mH$_{2}$S +
\phantom{2}mCO$_{\phantom{1}}$ $\rightarrow$ mCS$_{2}$ + mO + 2H$_2$ &
2.12$\times$10$^{-17}$\\ \phantom{2}mH$_{2}$S + \phantom{C}2mO$_{\phantom{1}}$
$\rightarrow$ mSO$_{2}$ + H$_2$ & 5.42$\times$10$^{-17}$\\ \hline \end{tabular}
\end{center} \end{table}

The rate constants, $k$ (s$^{-1}$), have been evaluated by the product
of the reaction cross section ($\sigma$ in cm$^{2}$) and the flux of
cosmic ions ($F_\mathrm{ISM}$ in cm$^{-2}$\,s$^{-1}$). To apply the
laboratory results to interstellar medium conditions both quantities
have been corrected using the following
assumptions:
\begin{enumerate} \item We derive the flux of cosmic ions
  in the approximation of effectively mono-energetic 1\,MeV protons
  \citep{Mennella03}: $F_\mathrm{ISM}$ =
  0.22\,ions\,cm$^{-2}$\,s$^{-1}$.  $F_\mathrm{ISM}$ must be regarded
  as an effective quantity: it represents the equivalent flux of
  1\,MeV protons which gives rise to the ionisation rate produced by
  the cosmic ray spectrum if 1\,MeV protons were the only source for
  ionisation.\\ \item Furthermore, we speculate that the cross section
  scales with the stopping power ($SP$, the energy loss per unit path
  length of impinging ions). According to the SRIM code by
  \citet{Ziegler09}, in the case of protons impinging on a CO:H$_{2}$S
  mixture, $SP$(200\,keV protons)= 2.7$\times SP$(1\,MeV
  protons). \end{enumerate}
In the case of hydrogen sulfide we define $\sigma_\mathrm{ISM}$ =
$\sigma$/2.7\,cm$^{2}$ and we derive a destruction rate of hydrogen
sulfide equal to 3.8$\times$10$^{-16}$\,s$^{-1}$. In the case of
species formed after irradiation we define $\sigma_\mathrm{ISM}$ =
$p$/2.7\,cm$^{2}$ (see Table~\ref{tab:detectmol} for $p$ values) and
the calculated formation rates for each species are given in
Table~\ref{tab:dissoh2s}.

In our code two-body grain-surface reactions are considered to be
bimolecular reactions occurring as they would in the gas phase,
meaning that the parameter $\alpha$ should be expressed in units of
volume (cm$^3$\,s$^{-1}$). Since the data from experiments are in
units of time, we have followed a procedure
\citep[see][]{Occhiogrosso11} for evaluating the rate coefficients for
the formation of S-bearing species in the icy mantle, {\it
  viz-\`a-viz}, we consider an excess of one of the reactants. The
rate of the reaction will therefore vary only with the concentration
of the second reactant. As the total amount of ice varies with time,
we calculate a value for the rate of each reaction that varies with
time. For instance, when we consider CO as the most abundant reactant
(this is the case of the second and the fourth reaction in
Table~\ref{tab:dissoh2s}), reaction rates cover six orders of
magnitude because of the wide interval spanned by the CO abundances
during the collapse phase of protostar formation.  We point out that
density and freeze-out rate indeed play a pivotal role in controlling
the trends of the molecular abundances and since both of these two
physical parameters only significantly change towards the end of the
collapse, we therefore refer only to the final abundance of CO in
order to scale our reaction rates.

\subsubsection{Laboratory investigations of solid OCS formation}
\label{subsec:labocs}

In the experiments at the Cosmic Dust Laboratory (Department of
Chemistry, University College London) reactants are co-deposited on to
a highly ordered pyrolytic graphite (HOPG) substrate. During the
dosing period, the substrate is held at a constant temperature,
typically in the range 12--100\,K. Once the reactants are deposited,
the material is allowed to cool to 12\,K. Following the deposition,
the sample is heated up to 200--300\,K (depending on the system
studied) during which time mass spectra are recorded. The result is a
histogram of ion intensity as a function of the ion mass-to-charge
($m$/$z$) ratio and the surface temperature. These experiments are
repeated at different temperatures to give the dependence of the
amount of product formed on the substrate during dosing. Finally, a
simple kinetic model is run to derive reaction barriers and Arrhenius
pre-exponential factors from the temperature profile, allowing rate
constants to be calculated. In particular, \citet*{Ward12} studied the
following route for the formation of solid
OCS: \begin{equation} \label{eq:ocssolidform} \rm mCS_{2} + mO
  \rightarrow mOCS + mS, \end{equation} with CS$_{2}$ and O that are
both adsorbed on the icy mantles. They found a rate constant of
1.24$\times$10$^{-20}$\,cm$^{2}$\,molecule$^{-1}$\,s$^{-1}$. As
mentioned in the previous section, since our code accounts for
reactions that occur in three dimensions (i.e. as gas phase
reactions), we need to transform the surface rate constants from
\citet{Ward12} into standard units, cm$^{3}$ s$^{-1}$ \citep[for more
  details on the theoretical assumptions in this transformation,
  see][]{Occhiogrosso12}. We therefore calculate a final value of the
rate constant as 6.4$\times$10$^{-23}$\,cm$^{3}$\,s$^{-1}$ at 20\,K
for the reaction between O and CS$_{2}$.

\subsection{Incorporation of new theoretical data into the model}

In order to investigate the form of sulfur once it freezes on to the grain
surface, we insert an extended chemistry including all the S-bearing species
mentioned in the previous Subsection into our gas-grain chemical network. In
addition to these reactions, we also insert two new paths for the formation of
carbonyl sulfide, OCS, as theoretically investigated by \citet{Adriaens10}.
Reactions are listed in Table~\ref{tab:ocspaths}.

\begin{table} \begin{center} \caption{Routes to OCS formation on a coronene
surface, from \citet{Adriaens10}. $\alpha$, $\beta$, $\gamma$ represent the
parameters for the rate coefficient in the modified Arrhenius equation. 
They have been adapted for a water-ice surface by Adriaens (2013, priv. comm.).
\label{tab:ocspaths}} \begin{tabular}{lccc} \hline Reaction & $\alpha$
[s$^{-1}$] & $\beta$ & $\gamma$ [K]\\ \hline CO + S\phantom{H} $\rightarrow$ OCS
& 1.66$\times$10$^{-11}$ & 0 & 1893\\ CO + HS $\rightarrow$ OCS + H &
1.66$\times$10$^{-11}$ & 0 & \phantom{1}831\\ \hline \end{tabular} \end{center}
\end{table} The rate parameters relative to each channel are also reported;
$\gamma$ represents the reaction barrier in units of Kelvin (K). Note that,
unlike \citet{Adriaens10}, we do not account for the formation of any 
adduct species and we do not distinguish among the  cis-trans geometries
of the reactants and the products.  The adsorption energies calculated by
\citet{Adriaens10} were smaller in value than those experimentally determined in
previous studies \citep{Piper84, Mattera80} due to the fact that the theory
considers a perfect coronene surface and neglects the weak physisorption
interactions observable in the presence of substrate defects. Every reaction
activation barrier given can therefore be seen as an upper limit to the
effective barrier.

In addition to incorporating a grain-surface reaction network for
sulfur-bearing species as a revision to the original version of the code, we
also update our gas-phase reaction network with new rate coefficients taken from
the KIDA database \citep{Wakelam12}. In particular, we insert or amend the
rate parameters  from the UMIST database for different OCS routes of
formation as described in \citet[][see Table~\ref{tab:gasphaseocs}]{Loison12}.
\begin{table*}
\begin{center} 
\caption{Gas phase paths of OCS formation and
destruction and their competitive routes. The revised $\alpha$ values are taken from
\citet{Loison12}. \label{tab:gasphaseocs}}
\begin{tabular}{lcc} \hline 
Reaction & New $\alpha$ values & Previous $\alpha$
values \\ & [cm$^{3}$mol$^{-1}$s$^{-1}$] & [cm$^{3}$mol$^{-1}$s$^{-1}$] \\
 \hline CH + SO\phantom{H} $\rightarrow$ OCS + H\phantom{CO} &
1.1$\times$10$^{-10}$ &  -- \\ CH + SO\phantom{H} $\rightarrow$ SH\phantom{O} +
CO\phantom{H} & 9.0$\times$10$^{-11}$ & -- \\ O\phantom{H} + HCS $\rightarrow$
OCS + H\phantom{CO} & 5.0$\times$10$^{-11}$ & 5.0$\times$10$^{-11}$\\
O\phantom{H} + HCS $\rightarrow$ SH\phantom{O} + CO\phantom{H} &
5.0$\times$10$^{-11}$ & -- \\ H\phantom{C} + HCS $\rightarrow$
H$_{2}$\phantom{O} + CS\phantom{N} & 1.5$\times$10$^{-10}$ &
4.0$\times$10$^{-9}$\\ S\phantom{H} + HCO $\rightarrow$ OCS + H\phantom{CO} &
8.0$\times$10$^{-11}$ & -- \\ S\phantom{H} + HCO $\rightarrow$ SH\phantom{O} +
CO\phantom{H} & 4.0$\times$10$^{-11}$ & 3.6$\times$10$^{-10}$\\ OH +
CS\phantom{H} $\rightarrow$ OCS + H & 1.7$\times$10$^{-10}$ &
9.4$\times$10$^{-14}$\\ OH + CS\phantom{H} $\rightarrow$ SH\phantom{O} +
CO\phantom{H} & 3.0$\times$10$^{-11}$ & -- \\ C\phantom{H} + OCS $\rightarrow$
CO\phantom{S} + CS\phantom{N} & 1.0$\times$10$^{-10}$ & 1.6$\times$10$^{-9}$\\
CH + OCS $\rightarrow$ CO\phantom{S} + CS\phantom{N} + H\phantom{H} &
4.0$\times$10$^{-10}$ & -- \\ CN + OCS $\rightarrow$ CO\phantom{S} + NCS &
1.0$\times$10$^{-10}$ & -- \\ \hline \end{tabular} 
%\footnotesize {$\alpha$ units: cm$^{3}$mol$^{-1}$s$^{-1}$} 
\end{center} \end{table*}

\section{Presentation of results} \label{sec:modgrid}

\subsection{Trends among the data} \label{subsec:analtrends}

We commence by setting the initial fractional abundance of gaseous sulfur ions
equal to 1.4$\times$10$^{-6}$ \citep*[as measured by][]{Sofia94}. We also fix
the physical parameters to values typical of high mass SFRs, listed in
Table~\ref{tab:input}. \begin{table} \begin{center} \caption{Physical parameters
for the model of a prototypical high-mass star. {\tt fr} gives an indication of
the freeze-out efficiency, {\tt maxt} is the maximum temperature reached in the
model, {\tt size} is the diameter of the hot core, {\tt dens} ({\tt df}) is the
(maximum) density reached in the collapse, at a time earlier than {\tt tfin}.
$\zeta$ represents the cosmic ray ionisation rate. \label{tab:input}}
\begin{tabular}{@{}lc} \hline Parameter & Value\\ \hline {\tt temp} [K] & 10\\
{\tt fr} [\%] & 98\\ {\tt maxt} [K] & 100\\ {\tt size} [pc] & 0.03\\ {\tt dens}
[cm$^{-3}$] & 2$\times$10$^{2}$\\ {\tt df} [cm$^{-3}$] & 1$\times$10$^{7}$\\
{\tt tfin} [yr] & 10$^{7}$\\ {\tt $\zeta$} [s$^{-1}$] & 1.3$\times$10$^{-17}$\\
\hline \end{tabular} \end{center} \end{table}

Since we aim to investigate the effect of the new surface reactions on the
abundances of the S-bearing species, we have to allow molecules to freeze on to
the grain. In particular, we consider two limiting cases where each species
freezes as itself (F1) or it hydrogenates (F2) and an intermediate case (F3)
where we hydrogenate half of the accreting S-bearing species. The grid of the
freeze-out pathways chosen is reported in Table~\ref{tab:combinations}. Note
that not all the sulfur-bearing species included in our model are shown; in
particular, we assume that H$_{2}$S, H$_{2}$CS, OCS, SO$_{2}$, NS, SO, CS$_{2}$
and H$_{2}$S$_{2}$ accrete on to the grains without hydrogenating.

\begin{table*} \begin{center} \caption{Grid of possible paths of freeze-out for
S-bearing molecules. \label{tab:combinations}}
\footnotesize
\begin{tabular}{@{}cllllll} \hline 
Model & S$\rightarrow$grains &
HS$\rightarrow$grains & HS$_{2}$$\rightarrow$grains & S$_{2}$$\rightarrow$grains
& CS$\rightarrow$grains & HCS$\rightarrow$grains\\ \hline \multirow{2}{*}{F1} &
100$\%$ mS & 100$\%$ mHS & 100$\%$ mHS$_{2}$ & 100$\%$ mS$_{2}$ & 100$\%$
mCS$_{\phantom{2}}$ & 100$\%$ mHCS$_{\phantom{2}}$\\ & \phantom{10}0$\%$
mH$_{2}$S & \phantom{10}0$\%$ mH$_{2}$S & \phantom{10}0$\%$ mH$_{2}$S$_{2}$ &
\phantom{10}0$\%$ mH$_{2}$S$_{\phantom{2}}$ & \phantom{10}0$\%$
mHCS\phantom{H$_{2}$} & \phantom{10}0$\%$ mH$_{2}$CS\\ \hline
\multirow{2}{*}{F2} & \phantom{10}0$\%$ mS & \phantom{10}0$\%$ mHS &
\phantom{10}0$\%$ mHS$_{2}$ & \phantom{10}0$\%$ mS$_{2}$ & \phantom{10}0$\%$
mCS$_{\phantom{2}}$ & \phantom{10}0$\%$ mHCS$_{\phantom{2}}$\\ & 100$\%$
mH$_{2}$S & 100$\%$ mH$_{2}$S & 100$\%$ mH$_{2}$S$_{2}$ & 100$\%$
mH$_{2}$S$_{\phantom{2}}$ & 100$\%$ mHCS\phantom{H$_{2}$} & 100$\%$ mH$_{2}$CS\\
\hline \multirow{2}{*}{F3} & \phantom{1}50$\%$ mS & \phantom{1}50$\%$ mHS &
\phantom{1}50$\%$ mHS$_{2}$ & \phantom{1}50$\%$ mS$_{2}$ & \phantom{1}50$\%$
mCS$_{\phantom{2}}$ & \phantom{1}50$\%$ mHCS$_{\phantom{2}}$\\ &
\phantom{1}50$\%$ mH$_{2}$S & \phantom{1}50$\%$ mH$_{2}$S & \phantom{1}50$\%$
mH$_{2}$S$_{2}$ & \phantom{1}50$\%$ mH$_{2}$S$_{\phantom{2}}$ &
\phantom{1}50$\%$ mHCS\phantom{H$_{2}$} & \phantom{1}50$\%$ mH$_{2}$CS\\ \hline
\end{tabular} \end{center} \end{table*}

In addition, we need to derive an estimate of the sulfur species in the
gas-phase, after their thermal evaporation from the grain surface, for
comparison to evolved hot cores. As investigated by \citet{Viti04}, species can
evaporate from the icy mantle in different bands of temperatures; in particular,
\citeauthor{Viti04} determined that H$_{2}$CS behaves as a H$_{2}$O-like
molecule that co-desorbs with the H$_{2}$O-ice when it starts to sublime from
the grain surface (at $\sim$100\,K). Moreover, the authors classified HCS, OCS,
H$_{2}$S, SO$_{2}$ as intermediate species since they showed two peaks (due to
volcano and co-desorption effects) in their Temperature Programmed Desorption
(TPD) traces. After further discussion (W.~A. Brown, priv. comm.), we have
revised the above desorption classification \citep{Viti04} as follows:
\begin{enumerate} \item  HS, H$_{2}$S$_{2}$, OCS, H$_{2}$S, SO$_{2}$, HCS, NS as
intermediate \item  H$_{2}$CS, SO as water-like \item  HS$_{2}$ as reactive
\item  S$_{2}$, CS$_{2}$, S-residue as refractory. \end{enumerate} No
sulfur-bearing species are classified as having a desorption behaviour that is
CO-like \citep[the initial desorption category of][]{Viti04}.

In order to assess the impact of our updates on the chemical network, we ran
some test models, comparing our original version of the code and those including
the new sets of reactions. Our findings are reported in
Table~\ref{tab:modelcomparison}. The changes in the ice composition are more
evident at the end of the collapse phase (Phase I), when temperatures are as low
as 10\,K. Note that we only show our results assuming the highest degree of
hydrogenation (F2) during the freeze-out of the molecules and we reserve a more
detailed investigation for later in the paper. \begin{table} \begin{center}
\caption{Fractional abundances (with respect to H$_{2}$) of solid sulfur-bearing
species obtained as outputs from our code before (OLD) and after (NEW) our
updates at the end of Phase I. The $m$ before the molecular formula stands for
{\it mantle}. \label{tab:modelcomparison}} \begin{tabular}{@{}lcc} \hline
Species & OLD & NEW \\ \hline mCS & 2.6$\times$10$^{-13}$ &
2.6$\times$10$^{-13}$\\ mOCS & 3.9$\times$10$^{-09}$ & 6.6$\times$10$^{-09}$\\
mHCS & 1.6$\times$10$^{-08}$ & 1.8$\times$10$^{-08}$\\ mH$_{2}$S &
1.4$\times$10$^{-06}$ & 1.4$\times$10$^{-06}$\\ mH$_{2}$CS &
5.8$\times$10$^{-09}$ & 5.7$\times$10$^{-09}$\\ mS & trace & trace \\ mS$_{2}$ &
5.8$\times$10$^{-13}$ & 2.4$\times$10$^{-13}$\\ mSO & 8.8$\times$10$^{-10}$ &
9.2$\times$10$^{-10}$\\ mSO$_{2}$ & 5.4$\times$10$^{-10}$ &
5.7$\times$10$^{-10}$\\ mNS & 9.9$\times$10$^{-12}$ & 1.1$\times$10$^{-11}$\\
mHS & trace & trace \\ mHS$_{2}$ & 3.2$\times$10$^{-14}$ &
3.7$\times$10$^{-13}$\\ mH$_{2}$S$_{2}$ & 1.1$\times$10$^{-12}$ &
2.2$\times$10$^{-10}$\\ mCS$_{2}$ & none & 8.7$\times$10$^{-15}$\\ mS-residue &
none & 5.4$\times$10$^{-09}$\\ \hline \end{tabular} \end{center} \end{table}

Scrutinising Table~\ref{tab:modelcomparison}, we immediately observe that the
most abundant sulfur species in the solid state is H$_{2}$S, which is
predictable since we are discussing the model with the highest degree of
hydrogenation. The abundance of OCS is enhanced in the new model by a factor of
$\sim$2; this is mostly due to the reaction of H$_2$S + CO on the grain surface,
with the rate from \citet{Garozzo10}. The reaction studied by \citet{Ward12}
contributes at the level of a few percent. The most interesting result is
definitely the amount of sulfur residue (S-residue) that we are now able to
produce on the grain. As we can see, in the updated version of the code,
S-residue is in fact the fifth most abundant ice-mantle S-bearing species. These
findings are extremely important, both because they match what has been already
experimentally determined by \citet{Jimenez-Escobar14} as an attempt to explain
the lack of sulfur in dense regions.  We now need to provide a better estimate
of this residue in order to understand how its presence might affect the total
amount of interstellar sulfur; we have decided not to speculate more about its
nature, since only experiments can give us information on the chemical structure
of these refractory molecules, but refer the interested reader to
\citet{Steudel03}.

\begin{figure*} \begin{center} \includegraphics[width=150mm]{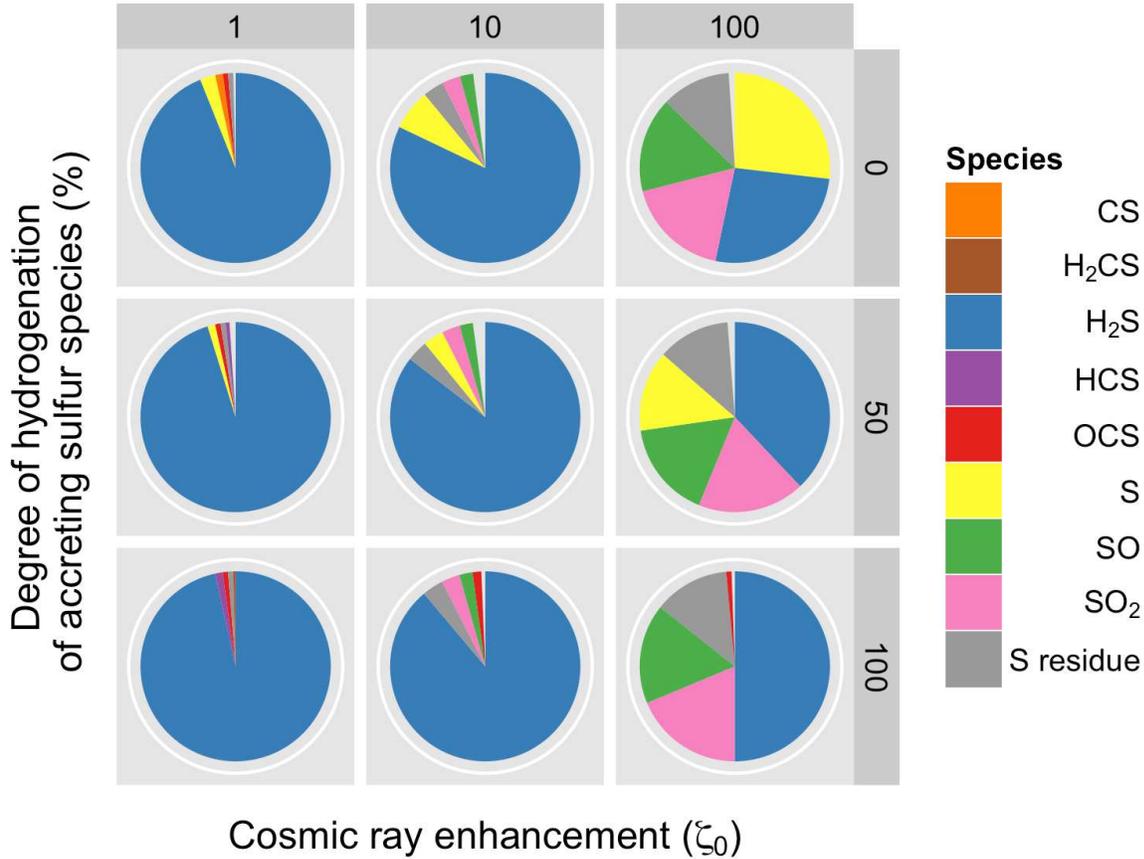}
\caption{The most abundant S-bearing species in the icy mantle at the end of the
collapse phase (Phase I). 0, 50 and 100 on the vertical axis refer to the percentage of
hydrogenation chosen (see text); 1, 10, 100 indicate a standard, an enhanced and
a super-enhanced cosmic ionisation rate, respectively. \label{fig:phase1}}
\end{center} \end{figure*}

\begin{figure*} \begin{center} \includegraphics[width=150mm]{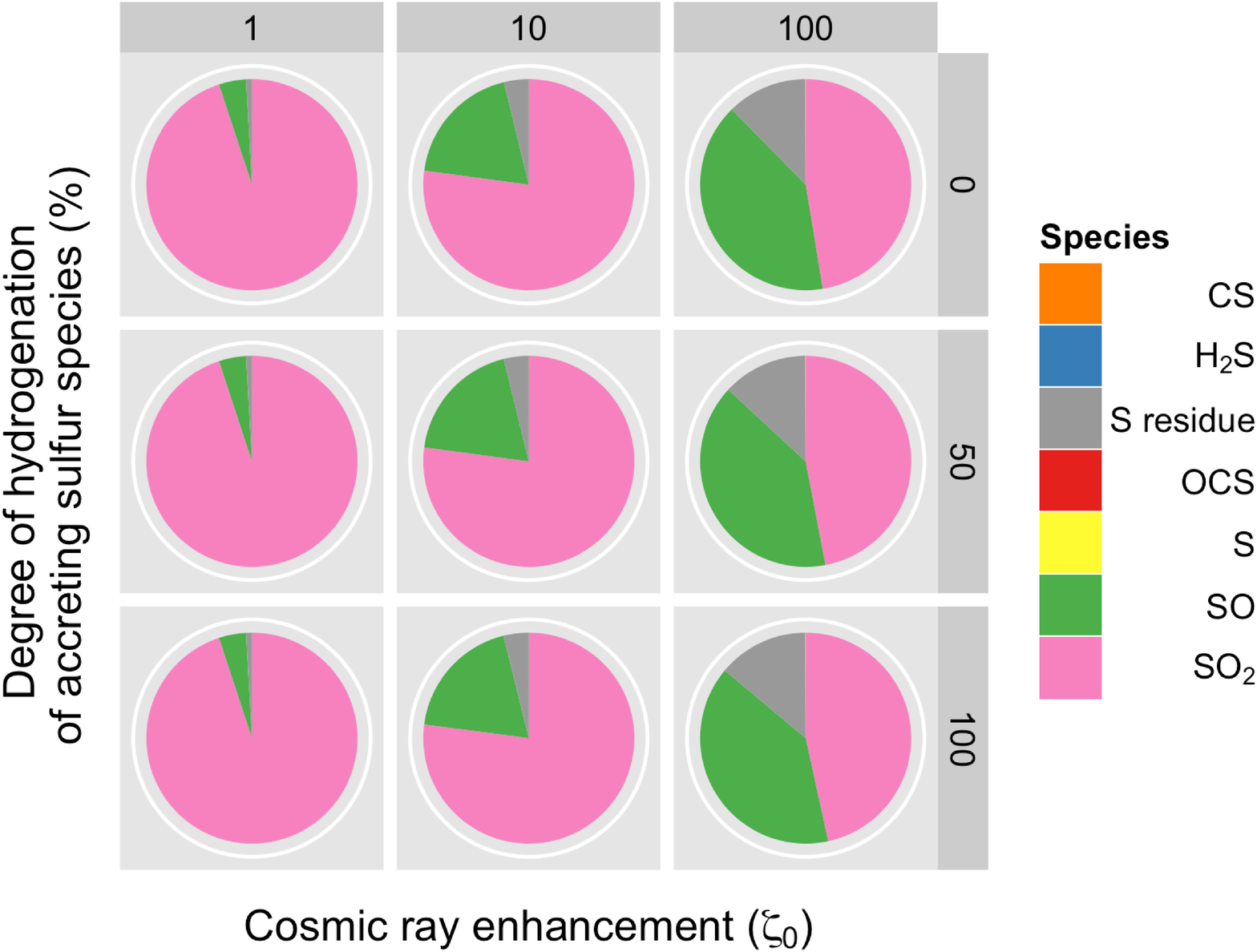}
\caption{The most abundant S-bearing species in the gas (excluding the residue,
which is solid) at the end of the desorption phase (Phase II). 0, 50 and 100 on the
vertical axis refer to the percentage of hydrogenation chosen (see text and
Table~\ref{tab:combinations}); 1, 10, 100 indicate a standard, an enhanced and a
super-enhanced cosmic ray ionisation rate, respectively. \label{fig:phase2}}
\end{center} \end{figure*}

Based on the computational work that has previously been performed in this area
(see Introduction) as well as on some simple test models, we have not
investigated changes in the temperature or in the density of our sources;
instead, we analyse how the calculated abundances vary with the cosmic-ray
ionisation rate by choosing a standard interstellar
($\zeta_0=$ 1.3$\times$10$^{-17}$\,s$^{-1}$), an enhanced (1.3$\times$10$^{-16}$\,s$^{-1}$)
and a super-enhanced (1.3$\times$10$^{-15}$\,s$^{-1}$) value for this parameter.
With these parameters we then repeat our sensitivity tests for each freeze-out
chemistry (F1, F2, F3; Table~\ref{tab:combinations}). Figure~\ref{fig:phase1}
and Fig.~\ref{fig:phase2} display the most abundant species found at the end of
the collapse phase (where temperatures are as low as 10\,K and only non-thermal
desorption can occur) and when the hot core is finally formed ($\geq$100\,K),
respectively. Note that while Fig.~\ref{fig:phase1} shows the different
molecules in the solid state (on the grain surface), in Fig.~\ref{fig:phase2} we
report the species in the gas phase, with the exception of the S-residue. 
Both figures show the phase where the bulk sulfur is located, and despite the
increase in $\zeta$, cosmic-ray induced desorption in not efficient enough in our models to
return a significant amount of sulfur to the gas phase in Phase I. Specifically,
in models with $\zeta = \zeta_0$, we find about 0.02\% of sulfur in the gas phase. In
models with $\zeta = 100\zeta_0$, this percentage rises to $\sim$0.5\% of the elemental
sulfur in the gas phase at the end of Phase I.

Phase I appears to be dominated by H$_{2}$S \citep[in agreement with
  the scheme of][]{Hatchell98}, where the abundance seems to be
influenced by the degree of ionisation as well as by the percentage of
hydrogenation. Starting the discussion of trends in the results by
considering the case with 0 per cent hydrogenation (top panels), going
from left to the right in Figure~\ref{fig:phase1}, the increase in the
$\zeta$ values (from the standard to the super-enhanced) leads to a
more efficient non-thermal desorption of H$_{2}$S, which therefore
decreases in the icy mantle in favour of other S-bearing species. The
immediate consequence is a greater amount of hydrogen sulfide in the
gas phase. The latter species then reacts with atomic ions (C$^+$,
S$^+$, H$^+$) at early times and molecular ions (H$_3$O$^+$, HCO$^+$)
at late times to liberate atomic sulfur, which goes on to react with
OH in order to produce SO and SO$_{2}$. These oxides are then frozen
back on to the grain.  Moreover, we highlight a predictable increase
in the S-residue abundance (since the rate of cosmic ray ionisation
directly depends on the $\zeta$ value) and an increasingly inefficient
formation of solid OCS, which is mainly produced on the grain by
reaction between H$_{2}$S and CO.  Moving to the bottom panels, the
chemical behaviour is altered by the higher level of hydrogenation. At
the beginning the situation is very similar to the one described
above, but once gaseous HS and S are formed, they will both freeze
back on to the grain as H$_{2}$S. Therefore, assuming a super-enhanced
cosmic ray ionisation rate, compared to the previous case, we now
observe a larger amount of H$_{2}$S, OCS and S-residue in the solid
state.

Phase II is dominated by SO$_{2}$ and the second most abundant species is SO.
This result is not surprising. If we look at the chain of reactions mentioned
above, we actually find that gaseous H$_{2}$S, S and HS will eventually lead to
the formation of SO and SO$_{2}$. The chemistries of these two species are
strongly related to each other: SO forms SO$_{2}$ by reaction with O or OH;
SO$_{2}$ produces SO when reacting with C. We notice immediately from
Fig.~\ref{fig:phase2} that the SO/SO$_2$ ratio is sensitive to the cosmic-ray
ionisation field, increasing with field strength. Finally, as we can see in
Fig.~\ref{fig:phase2}, an observable abundance of S-residue is still locked
inside the grain, even after the ice mantle has been completely desorbed.

\subsection{Comparison of theoretical results with recent observations}
\label{sec:obs}

\begin{figure} \begin{center} \includegraphics[width=85mm]{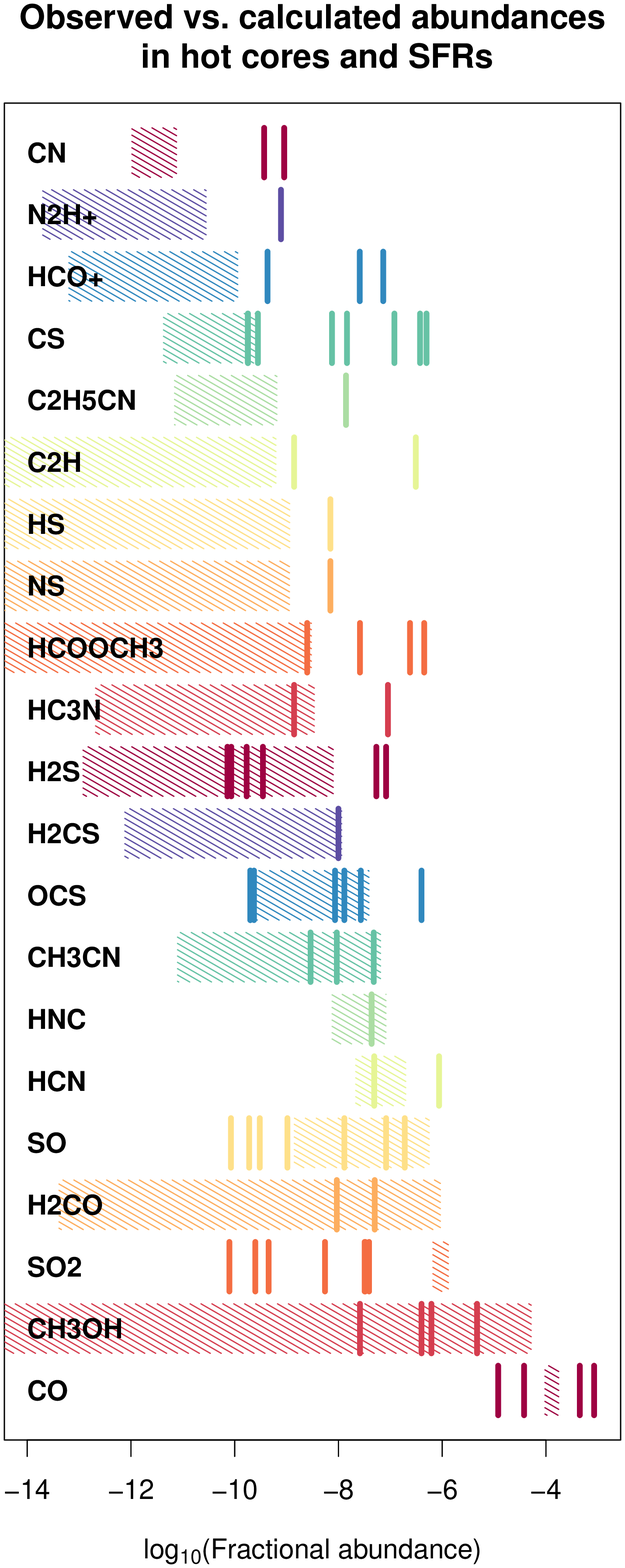}
\caption{Observed (solid coloured lines) and ranges of calculated abundances
from our models (hatched regions) for the species reported in a number of recent
observational papers \citep{Herpin09,Qin10,Neufeld12,Zernickel12,Xu13,Gerner14},
dealing with hot cores and high-mass SFRs. \label{fig:obsabunds}}
\end{center} \end{figure}

In order to verify the reliability of our calculations, we looked at the
observed abundances of 21 species in hot cores and high-mass SFRs reported in
the literature in the last five years. The reader should bear in mind that when
performing a comparison such as this, order of magnitude agreement between
observation and model results is excellent. Telescope beams often cover large
regions, resulting in average column densities and abundances (averaged over
both temperature and density variations), and since it is observationally very
hard to determine the evolutionary stage of a hot core, it becomes equally hard
to identify a \textquoteleft $t=0$\textquoteright\ to which to anchor the model
results. Despite these caveats, a comparison between observational and model
results is an important check of the model.

\citet{Herpin09} surveyed four hot cores using the IRAM 30\,m and CSO
telescopes, and detected lines of CS, OCS, H$_2$S, SO, SO$_2$ and their $^{34}$S
isotopologues. \citet{Qin10} used the SMA interferometer to observe G19.61-0.23,
detecting 17 molecules, including SO, SO$_2$, OCS and CS and various
isotopologues. Of these four molecules, only the SO abundance was determined
using more than one line. \citet{Neufeld12} discovered the mercapto radical (HS)
in the high-mass SFR, W49N, and also provided a measurement of the H$_2$S
abundance there. \citet{Zernickel12} performed a sub-mm survey of the high-mass
SFR NGC 6334I, detecting over 20 molecules in two hot cores. Several of these
molecules are sulfur-bearing: OCS, CS, SO, H$_2$S, SO$_2$ and H$_2$CS.
\citet{Xu13} also used the SMA interferometer to observe a massive SFR,
G20.08–0.14N, detecting 11 molecular species, including SO and SO$_2$.  Finally,
\citet{Gerner14} observed a large sample of 59 high-mass SFRs at various stages
of evolution, including 11 hot molecular cores. Sixteen molecules were detected,
of which four -- $^{13}$CS, SO, C$^{33}$S, OCS -- contain sulfur. Note that all
these cores span a range of masses, distances and evolutionary stages.

We compared fractional abundances calculated by our models at 10$^5$\,yr of
Phase II, for three values of the cosmic-ray ionisation rate in
Fig.~\ref{fig:obsabunds}. Variations in hydrogenation efficiency during Phase I
have little effect on the gas-phase abundances in Phase II, and thus we present
results from ``50 per cent hydrogenation'' models only. In general, the
agreement between calculated abundances and abundances derived from the
observations is very good: models and observational results are in agreement for
two thirds of the species, and in the some of the remaining third, disagreement
is less than an order of magnitude. There are issues with species which are
formed mostly by UV and CR dissociation: CN, N$_2$H$^+$, C$_2$H and HCO$^+$ are
under-estimated by our model. In the model, interstellar UV photons are assumed
to become extinguished in the outer core material.  Only cosmic-ray induced UV
photons play an active part in the chemistry of our simulations.  Hot cores, in
reality, may be significantly fragmented, allowing UV photons to penetrate more
deeply into the core than in a homogeneous medium. This may account for our
under-estimation of these species in our model. There is little evidence in
our galaxy for cosmic-ray ionisation rates higher than those tested in our
model \citep{Dalgarno06}.

Observations indicate that CS is the most abundant sulfur-bearing species,
however models struggle to produce enough CS to match observed amounts. This is
a recurrent problem in warm-temperature chemical models
\citep[e.g.,][]{Wakelam11}. This under-estimation may stem from a lack of
understanding of the warm-temperature chemistry of CS, incorrect rate
coefficients for its gas-phase reactions, shock passage through the observed
regions \citep{Viti01}, etc. Other sulfur-bearing species are observed to have
similar abundances to each other: OCS, H$_2$S, SO, SO$_2$, H$_2$CS and HS, all
have average fractional abundances in the range 7--70$\times$10$^{-9}$ in
Fig.~\ref{fig:obsabunds}.  Simulation results are within the same range for all
the above species with the exception of SO$_2$, which seems to be
over-estimated, and SO, which appears to be under-produced (the relationship
between these two species has already been discussed in detail -- see Sect.
~\ref{subsec:analtrends}).

\begin{table*} \begin{center} \caption{Calculated fractional abundances of
selected sulfur-bearing species as a function of cosmic-ray ionisation rate,
compared to their mean observed values towards various hot cores. Data are taken
from \citet{Herpin09,Qin10,Neufeld12,Zernickel12,Xu13} and \citet{Gerner14}.
\label{tab:finaltab}} \begin{tabular}{@{}lcccc} \hline Species & Standard Model
& Enhanced Model & Super-enhanced Model & Observed range\\ \hline SO$_{2}$ &
1.3$\times$10$^{-6}$  & 1.3$\times$10$^{-6}$  & 7.2$\times$10$^{-7}$  &
7.9$\times$10$^{-11}$--3.9$\times$10$^{-8}$\\ SO            &
4.9$\times$10$^{-9}$  & 1.5$\times$10$^{-9}$  & 6.1$\times$10$^{-7}$  &
8.5$\times$10$^{-11}$--1.9$\times$10$^{-7}$ \\ OCS           &
2.9$\times$10$^{-8}$  & 5.5$\times$10$^{-9}$  & 2.6$\times$10$^{-10}$ &
2.0$\times$10$^{-10}$--4.0$\times$10$^{-7}$\\ H$_{2}$CS     &
1.1$\times$10$^{-8}$  & 3.9$\times$10$^{-9}$   & trace                &
1.0$\times$10$^{-8}$\\ H$_{2}$S      & 8.5$\times$10$^{-9}$  & trace &
1.0$\times$10$^{-12}$ & 7.3$\times$10$^{-11}$--8.3$\times$10$^{-8}$\\ CS &
1.0$\times$10$^{-10}$ & 4.1$\times$10$^{-12}$  & 3.1$\times$10$^{-10}$ &
1.8$\times$10$^{-10}$--5.0$\times$10$^{-7}$\\ HS & 1.2$\times$10$^{-9}$  & trace
                & trace                 & 7.0$\times$10$^{-9}$    \\ NS         
  & 1.2$\times$10$^{-9}$  & trace & trace                 & 4.2$\times$10$^{-9}$
\\ HCS           & 5.5$\times$10$^{-11}$ & trace                 & trace        
        & $\ldots$ \\ S             & 3.2$\times$10$^{-11}$ &
1.9$\times$10$^{-11}$  & 3.5$\times$10$^{-10}$  & $\ldots$ \\ HS$_{2}$       &
4.0$\times$10$^{-10}$ & 2.3$\times$10$^{-10}$  & trace                & $\ldots$
\\ H$_{2}$S$_{2}$ & 1.7$\times$10$^{-9}$  & 6.8$\times$10$^{-10}$  & trace      
         & $\ldots$ \\ S$_{2}$       & 2.4$\times$10$^{-9}$  & trace            
    & trace & $\ldots$ \\ mS$_{2}$       & trace                &
1.5$\times$10$^{-12}$   & 7.9$\times$10$^{-12}$ & $\ldots$ \\ mCS$_{2}$      &
trace                & trace                  & trace                & $\ldots$
\\
S-residue & 5.5$\times$10$^{-9}$ & 2.3$\times$10$^{-8}$ & 8.5$\times$10$^{-8}$ &
$\ldots$ \\ \hline \end{tabular} \end{center} \end{table*}

In Table~\ref{tab:finaltab}, we present the abundances of sulfur-bearing species
for three values of $\zeta$, the cosmic-ray ionisation rate. Note that we also
include the abundances of species which have not been observed to date and the
amount of a potential residue. No ions are shown, since they are of low
abundance ($x$(ion)$\leq$10$^{-14}$), and do not contribute significantly to the
sulfur budget. As previously mentioned, apart from CS and SO$_2$, agreement is
good between calculated abundances of sulfur-bearing species and their observed
values. An interesting finding is that the abundance of H$_2$S$_2$ reaches
observable levels; therefore, the absence of its detection to date might be only
due to the fact that this molecule has a small dipole moment ($\sim$1.2\,D).

% Dipole moment from Maciel, Barreto et al. (2008), J Chem Phys, 129, 164302
% 
\begin{figure*} \begin{center} \includegraphics[width=150mm]{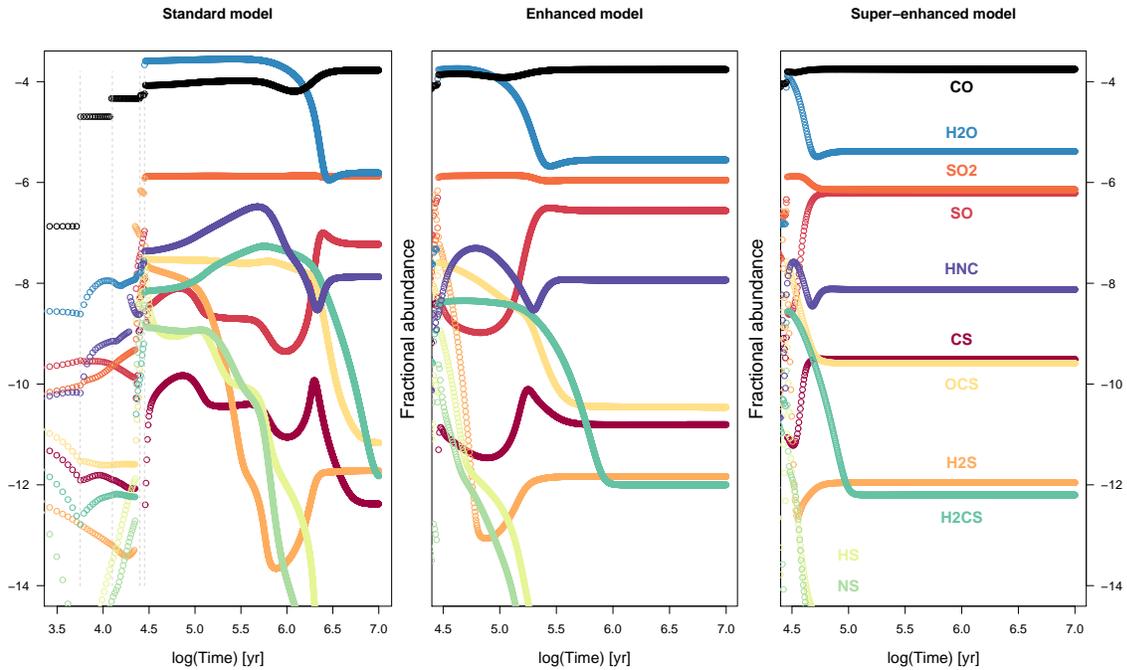}
\caption{Results from the warm-up phase of the model, showing the effect of the
cosmic-ray ionisation rate upon the fractional abundances of various species.
The left panel is extended slightly to show the desorption events \citep[those
of the {\it (i)} pure species on the surface of the ice, {\it (ii)} monolayer on
H$_2$O ice, {\it (iii)} volcano desorption, and {\it (iv)} co-desorption with
H$_2$O. See][for a full description]{Viti04}. The right panel shows species
labels. This figure highlights the issues with comparing model results at a
certain time to observational values, since fractional abundances can vary by
several orders of magnitude.\label{fig:phaseIIev}} \end{center} \end{figure*}

From Table~\ref{tab:finaltab} it is evident that the abundances of
most sulfur-bearing species are influenced by $\zeta$, with some
molecules only being significantly present in standard $\zeta$
environments (HS, NS, HCS). In order to make these changes more
evident, we plot the fractional abundance of the S-bearing species as
a function of the time during the desorption phase
(Fig.~\ref{fig:phaseIIev}).  In addition, we show the variation of
H$_{2}$O, HNC and CO, as standards for our models. In general,
abundant molecules, such as SO and OCS, are present at observational
levels at all three strengths of $\zeta$ and may be good tracers of
$\zeta$ if their observed abundances can be tightly
constrained. SO$_2$ is also abundant at high levels throughout the
tests, but does not vary much in abundance.

A further consideration evident in Fig.~\ref{fig:phaseIIev} is that the
abundances of some species vary considerably with time, and as stated in the
introduction to this section, this makes comparisons with observations
challenging. For instance, gas-phase H$_2$S abundances drop in all scenarios by
almost six orders of magnitude, but on different time-scales, and H$_2$CS
abundances by almost four orders of magnitude. Other species vary in abundance
by an order of magnitude or so, whereas species like SO$_2$ are present at their
solid-phase abundances, i.e., their abundances are unaltered by gas-phase
chemistry.

If we now consider the refractory sulfur-bearing species, S$_2$, CS$_2$ and the
sulfur residue, it is evident that an appreciable amount (up to 6 per cent) of
sulfur in our model is locked in form of this residue.  It is hard to speculate
about the chemical nature of this residue; recent experiments
\citep{Jimenez-Escobar14} seem to attribute its existence to the formation of
sulfur polymers, such as S$_{8}$. There may, of course, be several ways to sink
sulfur into this refractory residue, and we only consider a single route here,
led by the experimental evidence.

Finally, we are able to produce detectable abundances of species such as
H$_{2}$S and CS but only on the grain surface (Table~\ref{tab:modelcomparison}).
This means that, while there is a good understanding of the mantle sulfur
chemistry, the rates of some gas-phase reactions need to be refined; in
particular, the key step seems to be the high efficiency of SO and SO$_{2}$
formation from S and HS as reactants.

\section{Conclusions} \label{sec:conc}

In summary, we have extended the {\sc ucl\_chem} chemical model to include new
experimental and theoretical results on sulfur chemistry; in particular, we have
inserted laboratory data from experiments where H$_{2}$S ice is processed by
protons which simulate cosmic ray impacts on the grain surface.
\citet{Garozzo10} detected in their icy mantle analogues the formation of
CS$_{2}$, which has not been observationally detected in the gas-phase ISM yet.
In our modelling, we find that CS$_{2}$ was only present in small amounts in icy
mantles. Following on from the experiment by \citeauthor{Garozzo10},
\citet{Ward12} studied the production of solid OCS from CS$_{2}$ and O ice. We
therefore also added the latter reaction to our chemical network. OCS abundances
were enhanced by a factor of $\sim$2, driven mainly by the
\citeauthor{Garozzo10} reaction, mH$_2$S + mCO $\longrightarrow$ mOCS +
H$_2$, and only partially by the reaction from \citet{Ward12}.  Finally, we also
revised our gas-phase reactions based on recent papers by \citet{Loison12} and
\citet{Adriaens10}. We looked at the influence of the new reaction network on
the fractional abundances of selected S-bearing species by comparing the results
with the original version of our code. Abundances of species such as HS$_{2}$
and H$_{2}$S$_{2}$ increased by large factors in the solid phase via very
efficient hydrogenation of accreting species, but not significantly enough to
make them major carriers of sulfur. Our main result was that a residue, left on
the grain surface after ice desorption, could harbour a large amount of sulfur
(10$^{-8}$), in line with recent experiments \citep{Jimenez-Escobar14}.

A comparison with astronomical observations was also carried out. A good
agreement is obtained although we find it difficult to reconcile theoretical
estimates with observations of some of the most abundant species such as
H$_{2}$S, CS and OCS. Our results show that at early stages during the collapse
phase of star-formation, H$_{2}$S is the predominant sulfur-bearing species in
the icy mantles. As the evolution of a pre-stellar core continues SO$_{2}$ and
SO are efficiently formed, in agreement with earlier work. In this respect, the
pivotal processes appear to be the gas phase production of SO and SO$_{2}$ from
S and HS as reactants. Another key factor to be considered is the presence of a
sulfur residue on the grain surface, which may affect the observability of some
species in the gas-phase. This present paper is therefore an important step
forward in understanding the sulfur chemistry in regions of star formation as we
constrain the problem to two main factors: the gaseous interactions between S or
HS and O or OH to form SO and SO$_{2}$ and the existence of a sulfur residue on
the grain surface.

\section*{Acknowledgements} The research leading to these results has received
funding from the (European Union's) Seventh Framework Program [FP7/2007-2013]
under grant agreement n$^{\circ}$ 238258. This work was partly supported by the
Italian Ministero dell'Istruzione, Universit\`a e Ricerca (MIUR) through the
grant {\it Progetti Premiali 2012 - iALMA}. The research of Z.~K. has been
supported by VEGA -- The Slovack Agency for Science, Grant No. 2/0032/14.

\bibliographystyle{mn2e}
\bibliography{sulfur_residue_cleaned}
\label{lastpage}

\end{document}